\documentclass[pra,twocolumn]{revtex4-2}
\usepackage{amsmath}
\usepackage{amssymb}
\usepackage{graphicx}
\usepackage{braket}
\usepackage{stmaryrd}
\usepackage{hyperref}
\usepackage[usenames,dvipsnames]{xcolor}
\usepackage{dsfont}

\hypersetup{
    colorlinks,
    citecolor=blue,
    linkcolor=blue,
    urlcolor=blue
}

\begin{document}
\title{Speed limits and thermodynamic uncertainty relations for quantum systems with non-Hermitian Hamiltonian}

\author{Tomohiro Nishiyama}
\email{htam0ybboh@gmail.com}
\affiliation{Independent Researcher, Tokyo 206-0003, Japan}

\author{Yoshihiko Hasegawa}
\email{hasegawa@biom.t.u-tokyo.ac.jp}
\affiliation{Department of Information and Communication Engineering, Graduate
School of Information Science and Technology, The University of Tokyo,
Tokyo 113-8656, Japan}
\date{\today}

\begin{abstract}

Non-Hermitian Hamiltonians play a crucial role in describing open quantum systems and nonequilibrium dynamics. In this paper, we derive trade-off relations for systems governed by non-Hermitian Hamiltonians, focusing on the Margolus-Levitin-type  and Mandelstam-Tamm-type  bounds, which are originally derived  as quantum speed limits  in isolated quantum dynamics.
While the quantum speed limit for the Mandelstam-Tamm bound in general non-Hermitian systems was derived in the literature, we obtain a Mandelstam-Tamm quantum speed limit for continuous measurement using the continuous matrix product state formalism.
Moreover, we derive a Margolus-Levitin quantum speed limit in the non-Hermitian setting. 
We derive additional bounds on the ratio of the standard deviation to the mean of an observable, which take the same form as the thermodynamic uncertainty relation.
As an example, we apply these bounds to the continuous measurement formalism in open quantum dynamics, where the dynamics is described by discontinuous jumps and smooth evolution induced by the non-Hermitian Hamiltonian. Our work provides a unified perspective on the quantum speed limit and thermodynamic uncertainty relations in open quantum dynamics from the viewpoint of the non-Hermitian Hamiltonian, extending the results of previous studies.

\end{abstract}

\maketitle
\section{Introduction}

Non-Hermitian Hamiltonians are pivotal in various areas of quantum physics \cite{Ashida:2020:NonHermiteReview}, notably in investigating open quantum systems and analyzing dynamics that are not in equilibrium.
When a quantum system conserves its norm, its time evolution is described by a unitary transformation, which, in turn, implies that the generator associated with this unitary transformation is a Hermitian  operator.
A quantum system that interacts with a measuring device or its surrounding environment undergoes open quantum dynamics. 
For an open system which exchanges energy or particles, the state vector does not necessarily preserve the norm whose dynamics admits non-Hermitian Hamiltonian.
Typical examples of a non-Hermitian Hamiltonian include the Feshbach projection and quantum trajectory approaches.
The Feshbach projection approach \cite{Feshbach:1958:Nuclear,Feshbach:1958:Nuclear2} partitions the Hilbert space of a quantum system into two subspaces, a model space and its complementary space. Then, the approach projects the full Hamiltonian operator onto the model space, which yields an effective Hamiltonian describing the dynamics in the model space.
In the quantum trajectory approach \cite{Molmer:1993:MonteCarlo,Carmichael:2009:QuantumTrajLecture}, the system is coupled to an environment that undergoes continuous measurement (see Refs.~\cite{Daley:2014:QJReview,Landi:2023:CurFlucReviewPRXQ} for reviews). 
Given the measurement record, the state of the system can be described by a stochastic Schr\"odinger equation, which, on average, reduces to the Lindblad equation. 
In the quantum trajectory, the system experiences both abrupt, discontinuous jumps and smooth, continuous evolution.
The former is induced by the jump operators, whereas the latter smooth evolution is governed by the effective non-Hermitian Hamiltonian. 

In this paper, we derive trade-off relations for systems driven by non-Hermitian Hamiltonians. In particular, we focus on the Margolus-Levitin-type  \cite{Margolus:1998:QSL} and the Mandelstam-Tamm-type 
bounds \cite{Mandelstam:1945:QSL},
where lower bounds comprise the mean and the standard deviation, respectively, of physical quantities. 
These relations were originally derived in the context of quantum speed limits \cite{Mandelstam:1945:QSL,Margolus:1998:QSL,Deffner:2010:GenClausius,Taddei:2013:QSL,DelCampo:2013:OpenQSL,Deffner:2013:DrivenQSL,Pires:2016:GQSL} (see \cite{Deffner:2017:QSLReview} for a review).
Starting from an initial state, let $\tau$ be the time required for the system to become orthogonal to the initial state. 
The Margolus-Levitin and Mandelstam-Tamm quantum speed limits
provide lower bounds on $\tau$, given by the inverse of the average and the standard deviation of the Hamiltonian, respectively.
Here, we consider these two bounds for the case of non-Hermitian Hamiltonians, and further derive bounds on the ratio of the standard deviation to the average of an observable.
Such bounds on the ratio take the same form as the thermodynamic uncertainty relation \cite{Barato:2015:UncRel,Gingrich:2016:TUP,Carollo:2019:QuantumLDP,Hasegawa:2020:QTURPRL,Hasegawa:2020:TUROQS}.
Quantum speed limits for the non-Hermitian Hamiltonian were considered in Refs.~\cite{Uzdin:2012:NonHermitianSL,PhysRevA.104.052620,Sun:2019:GaugeInvariantQSL,Thakuria:2024:GQSL}. 
References~\cite{Uzdin:2012:NonHermitianSL,PhysRevA.104.052620} derived 
the Mandelstam-Tamm quantum speed limit for the non-Hermitian Hamiltonian. 
In Ref.~\cite{Thakuria:2024:GQSL}, the quantum speed limit for non-Hermitian systems was derived from the framework of the geometric quantum speed limit. Specifically, it defines the normalized state vectors and applies the Mandelstam-Tamm bound to the state to derive the bound.
References~\cite{Giovannetti:2003:QuantumLimits, Hoernedal:2023:MLQSL} obtained Margolus-Levitin-type speed limits given an arbitrary value of the fidelity between the initial and final states. 
In the present study, we derive a Margolus-Levitin quantum speed limit for non-Hermitian dynamics under the commutative assumption [cf. Eq.~\eqref{eq:commutative_condition}].
Despite the commutative assumption, our trade-off relations have the advantage of representing the cost term by the average of physical quantities.
Regarding uncertainty relations in non-Hermitian settings, Ref.~\cite{Pati:2015:NonHermiteWeakValue} derived a generalization of the Robertson uncertainty relation \cite{Robertson:1929:UncRel} in a non-Hermitian framework.
Specifically, Ref.~\cite{Pati:2015:NonHermiteWeakValue} defined an experimental protocol to evaluate the expectation of non-Hermitian operators. 
Besides quantum speed limits, we additionally derive thermodynamic uncertainty relations for both Margolus-Levitin-type and Mandelstam-Tamm-type bounds. 
In recent years, attempts have been made to understand the speed limit and thermodynamic uncertainty relation in a unified way \cite{Vo:2020:TURCSLPRE,Hasegawa:2023:BulkBoundaryBoundNC,Kwon:2023:XTUR,Nishiyama:2024:OpenQuantumRURJPA}.
As an example of the bounds, we apply the obtained bounds to the continuous measurement formalism in open quantum dynamics, where the dynamics is described by the discontinous jumps and smooth dynamics induced by a non-Hermitian Hamiltonian. 
Here, we provide a unified perspective on the quantum speed limit and thermodynamic uncertainty relations in open quantum dynamics from the viewpoint of the non-Hermitian Hamiltonian.

\begin{table*}
    \centering
\begin{tabular}{|c|c|c|}
\hline 
 & Margolus-Levitin type & Mandelstam-Tamm type\tabularnewline
\hline 
\hline 
Lower bound & ${\displaystyle 
\Lambda_{\mathrm{ML}}(0,\tau)=\frac{e^{-\langle\Gamma\rangle(0)\tau}-\tau\left(\langle H\rangle(0)-E_{g}\right)}{\||\psi(\tau)\rangle\|}
}$ & ${\displaystyle \Lambda_{\mathrm{MT}}\left(\tau_{1},\tau_{2}\right)=\cos\left[\int_{\tau_{1}}^{\tau_{2}}dt\Delta\mathcal{H}(t)\right]}$\tabularnewline
\hline 
$\begin{array}{c}
\text{Quantum}\\
\text{speed}\\
\text{limit}
\end{array}$ & $\displaystyle 
1+\frac{\tau\left(\braket{H}(0)-E_{g}\right)-e^{-\braket{\Gamma}(0)\tau}}{\|\ket{\psi(\tau)}\|}\geq 2\sin^{2}\left(\frac{\mathcal{L}_{D}(\widetilde{\rho}(0),\widetilde{\rho}(\tau))}{2}\right)
$ & ${\displaystyle \int_{\tau_{1}}^{\tau_{2}}dt\Delta\mathcal{H}(t)\geq\mathcal{L}_{D}\left(\widetilde{\rho}\left(\tau_{1}\right),\widetilde{\rho}\left(\tau_{2}\right)\right)}$\tabularnewline
\hline 
$\begin{array}{c}
\text{Thermodynamic}\\
\text{uncertainty}\\
\text{relation}
\end{array}$ & $\displaystyle 
\frac{\|\ket{\psi(\tau)}\|^2}{\left[e^{-\tau\braket{\Gamma}(0)}-\tau\left(\braket{H}(0)-E_{g}\right)\right]^{2}}-1\geq\left(\frac{\langle\mathcal{C}\rangle(\tau)-\langle\mathcal{C}\rangle(0)}{\Delta\mathcal{C}(\tau)+\Delta\mathcal{C}(0)}\right)^{2}
$ & $\displaystyle \left[\tan\left(\int_{\tau_1}^{\tau_2} dt \Delta \mathcal{H}(t) \right)\right]^2\geq \left(\frac{\langle \mathcal{C}\rangle(\tau_2)-\langle \mathcal{C}\rangle(\tau_1)}{\Delta\mathcal{C}(\tau_2)+\Delta\mathcal{C}(\tau_1)}\right)^2.$\tabularnewline
\hline 
\end{tabular}
    \caption{Summary of results. 
Lower bound $\Lambda(\tau_1,\tau_2)$, quantum speed limit, and thermodynamic uncertainty relation for Margolus-Levitin and Mandelstam-Tamm type tradeoff relations. 
$\mathcal{L}_D(\rho_1,\rho_2)$ is the Bures angle for density operators $\rho_1$ and $\rho_2$, and $\mathcal{C}$ is an Hermitian operator.
}
    \label{tab:results_summary}
\end{table*}

\section{Methods\label{sec:methods}}

\subsection{Non-Hermitian time evolution}

Consider the non-Hermitian Hemiltonian $\mathcal{H}$.
In general, $\mathcal{H}$ can be decomposed into
\begin{align}
    \mathcal{H}(t) = H(t) - i\Gamma(t),
    \label{eq:nonHermitianHamiltonian_decompose}
\end{align}
where $H(t)$ and $\Gamma(t)$ are Hermitian operators. 
In the Feshbach or quantum trajectory approach, 
it is guaranteed that $\Gamma(t)$ is semipositive definite \cite{Ashida:2020:NonHermiteReview}.  
Consider a state vector $\ket{\psi}$, whose time evolution is governed by the non-Hermitian Hamiltonian $\mathcal{H}(t)$:
\begin{align}
    i\frac{d}{dt}\ket{\psi(t)}=\mathcal{H}(t)\ket{\psi(t)}.
    \label{eq:Schrodinger1}
\end{align}
Throughout this paper, we adopt the convention of setting $\hbar =1$. 
Clearly, Eq.~\eqref{eq:Schrodinger1} is a generalization of the Schro\"odinger equation for the non-Hermitian case. 
It is possible to consider an initial mixed-state case.
 When the quantum state is density operator $\rho(t)$, 
Eq.~\eqref{eq:Schrodinger1} can be generalized to
\begin{align}
     \frac{d\rho}{dt}=-i(\mathcal{H}(t)\rho(t)-\rho(t)\mathcal{H}^{\dagger}(t)).
     \label{eq:nonHermitian_Heisenberg_eq}
 \end{align}
 Equation~\eqref{eq:nonHermitian_Heisenberg_eq} reduces to the Heisenberg equation when $\mathcal{H}(t)$ is Hermitian. 
Executing a purification process allows us to treat mixed states similarly to pure states.
Suppose that the initial state $\rho(0)$ satisfies $\mathrm{Tr}_S[\rho(0)] = 1$, where $\mathrm{Tr}_{S}$ is the trace with respect to the system, and has the following eigendecomposition:
\begin{align}
    \rho(0)=\sum_{l}p_{l}(0)\ket{\psi_{l}(0)}\bra{\psi_{l}(0)},
    \label{eq:rho_decompose}
\end{align}
where $p_l(0)$ is the eigenvalues satisfying $p_l(0)\ge0$ and $\sum_l p_l(0)=1$, and $\ket{\psi_l(0)}$ is the corresponding eigenvector. 
Then we can purify $\rho(0)$ as follows:
 \begin{align}
    \ket{\psi(0)}=\sum_{l}\sqrt{p_{l}(0)}\ket{\psi_{l}(0)}\otimes\ket{a_{l}},
    \label{eq:def_tilde_psi_main}
\end{align}
where  $\{\ket{a_{l}}\}$ are orthonormal basis in the ancilla.   
Note that the trace of the initial density operator is unity, and thus $\braket{\psi_l(0)|\psi_l(0)} = 1$.  
The time evolution of the purified state in Eq.~\eqref{eq:def_tilde_psi_main} is given by
\begin{align}
    i\frac{d}{dt}\ket{\psi(t)}=(\mathcal{H}(t)\otimes\mathbb{I}_{A})\ket{\psi(t)},
    \label{eq:purified_Schrodinger_eq}
\end{align}
where $\mathbb{I}_A$ is the identity operator on the ancilla (other identity operators are defined analogously). 
 Note that when the state is pure, there is no need to use the ancilla. 
In later calculations of expected values, we will leave out the $\mathbb{I}_A$ when it is evident.
It is easy to check
\begin{align}
    \rho(t)&=\mathrm{Tr}_{A}[\ket{\psi(t)}\bra{\psi(t)}]\nonumber\\&=\mathbb{T}e^{-i\int\mathcal{H}(t)dt}\rho(0)\overline{\mathbb{T}}e^{i\int\mathcal{H}^{\dagger}(t)dt},
    \label{eq:rhot_from_purification}
\end{align}
where $\mathbb{T}$ and $\overline{\mathbb{T}}$ denote time-ordering and anti time-ordering operators, respectively, and 
$\mathrm{Tr}_A$ is the partial trace with respect to the ancilla. 
Since $\mathcal{H}(t)$ is non-Hermitian, $\ket{\psi(t)}$ does not preserve the norm. 
Therefore, we introduce the normalized vector as follows:
\begin{align}
    \label{eq:def_norm_vec_main}
    \ket{\widetilde\psi(t)}&\equiv \frac{\ket{\psi(t)}}{\|\ket{\psi(t)}\|}, \\
    \label{eq:def_norm_density_main}
    \widetilde{\rho}(t)&\equiv \frac{\sum_l p_l \ket{\psi_l(t)}\bra{\psi_l(t)}}{\|\ket{\psi(t)}\|^2}=\mathrm{Tr}_{A}[\ket{\widetilde\psi(t)}\bra{\widetilde\psi(t)}],
\end{align}
where $\|\ket{\psi(t)}\|\equiv \sqrt{\braket{\psi(t)|\psi(t)}}$.

\subsection{Quantum speed limit}

The quantum speed limit \cite{Deffner:2017:QSLReview} sets an upper bound on the speed at which a quantum system can change. 
The original formulation due to Refs.~\cite{Mandelstam:1945:QSL,Margolus:1998:QSL} suggests that any isolated quantum system moving between two orthogonal states requires a minimum amount of time specified by the statistics of the Hamiltonian.
The quantum speed limit can be broadly divided into two categories: Mandelstam-Tamm \cite{Mandelstam:1945:QSL} and Margolus-Levitin \cite{Margolus:1998:QSL} type bounds.
In the Mandelstam-Tamm type bound, the minimum time is inversely proportional to the standard deviation of the Hamiltonian.
Suppose that the system is isolated and has a time-independent Hamiltonian $H$.
Then, the Mandelstam-Tamm bound states
\begin{align}
    \tau\geq\frac{\pi}{2\Delta H},
    \label{eq:MT_bound_def}
\end{align}
where $\Delta H\equiv\sqrt{\braket{H^{2}}-\braket{H}^{2}}$ is the standard deviation and 
$\braket{H}$ is the expectation of $H$. 
On the other hand, in the Margolus-Levitin bound, the minimum time is inversely proportional to the average of the Hamiltonian:
\begin{align}
    \tau\geq\frac{\pi}{2(\braket{H}-E_{g})}.
    \label{eq:ML_bound_def}
\end{align}
where $E_g\in\mathbb{R}$ is the minimum eigenvalue of $H$ (the ground state energy). 
$E_g$ is introduced to determine the reference point of the energy, since the expectation of the Hamiltonian can be arbitrary large while keeping the dynamics unchanged. 
The initial derivation of Eq.~\eqref{eq:ML_bound_def}, as presented in Ref.~\cite{Margolus:1998:QSL}, made use of a trigonometric inequality. 
Using a geometric approach, Ref.~\cite{Zwierz:2012:CommentOnGQSL} derived the slightly weaker bound:
\begin{align}
    \tau\geq\frac{1}{\braket{H}-E_{g}}.
    \label{eq:weaker_ML_bound_def}
\end{align}
The original derivation for Eq.~\eqref{eq:ML_bound_def} provides a tighter relation, but the geometric approach by \cite{Jones:2010:QSL,Zwierz:2012:CommentOnGQSL} offers better intuition and clarity.
A major distinction between the Mandelstam-Tamm and Margolus-Levitin quantum speed limits, given by Eqs.~\eqref{eq:MT_bound_def} and \eqref{eq:ML_bound_def}, respectively, lies in the fact that the Mandelstam-Tamm bound employed the standard deviation of the Hamiltonian, whereas the Margolus-Levitin bound utilizes the average of the Hamiltonian.
In the following, trade-off relations, such as quantum speed limits and thermodynamic uncertainty relations, are derived for quantum systems generated by non-Hermitian Hamiltonians.
Relations based on the average are called \textit{Margolus-Levitin type}, whereas those based on the standard deviation are referred to as \textit{Mandelstam-Tamm type}.

\subsection{Fidelity and tradeoff relations\label{sec:fidelity_tradeoff}}

In this manuscript, we derive both Mandelstam-Tamm and Margolus-Levitin type tradeoff relations, where the fidelity is crucial. 
Once the bound for the fidelity is obtained, we can derive quantum speed limits and thermodynamic uncertainty relations. 
In this section, we offer a comprehensive overview of the current study. 
Consider the fidelity $\left|\braket{\widetilde{\psi}(\tau_2)|\widetilde{\psi}(\tau_1)}\right|$, 
where $\ket{\widetilde{\psi}(t)}$ is the normalized vector defined in Eq.~\eqref{eq:def_norm_vec_main}. 
Suppose that the fidelity has a lower bound:
\begin{align}
    \left|\braket{\widetilde{\psi}(\tau_{2})|\widetilde{\psi}(\tau_{1})}\right|&\ge\Lambda(\tau_{1},\tau_{2}),
    \label{eq:illustrative_bound}
\end{align}
where $\Lambda(\tau_1,\tau_2)$ is a lower bound comprising the operators determined by the dynamics (e.g., Hamiltonian, jump operators, etc.).
Later, we will relate $\Lambda(\tau_1,\tau_2)$ with the physical quantities.

Let us derive quantum speed limits from Eq.~\eqref{eq:illustrative_bound}. 
In quantum speed limits, the distances between two quantum states are of our interest. 
Among distance measures, the Bures angle is widely employed in the literature \cite{Nielsen:2011:QuantumInfoBook}:
\begin{align}
    \mathcal{L}_D(\rho_1, \rho_2)\equiv \arccos\left[\sqrt{\mathrm{Fid}(\rho_1, \rho_2)}\right],
    \label{eq:L_D_def}
\end{align}
where $\mathrm{Fid}(\rho_1, \rho_2)$ is the quantum fidelity: 
\begin{align}
    \mathrm{Fid}(\rho_{1},\rho_{2})\equiv\left(\mathrm{Tr}\left[\sqrt{\sqrt{\rho_{1}}\rho_{2}\sqrt{\rho_{1}}}\right]\right)^{2}.
    \label{eq:fidelity_def}
\end{align}
Note that, for pure states, the fidelity becomes $\mathrm{Fid}(\ket{\psi(t_{1})},\ket{\psi(t_{2})})=|\braket{\psi(t_{2})|\psi(t_{1})}|^{2}$.
A notable property of the Bures angle is that it satisfies the monotonicity under completely-positive and trace preserving map (CPTP). 
Specifically, let $\mathcal{E}(\bullet)$ be a CPTP. Then the Bures angle satisfies
\begin{align}
    \mathcal{L}_{D}\left(\rho_{1},\rho_{2}\right)\ge\mathcal{L}_{D}\left(\mathcal{E}(\rho_{1}),\mathcal{E}(\rho_{2})\right).
    \label{eq:Bures_monotonicity}
\end{align}
Using the monotonicity of the fidelity, we can derive the quantum speed limit.
Suppose that $0 \le \Lambda(\tau_1,\tau_2) \le 1$.
Then the following relation holds:
\begin{align}
    \arccos\Lambda(\tau_{1},\tau_{2})&\ge\arccos\left|\braket{\widetilde{\psi}(\tau_{2})|\widetilde{\psi}(\tau_{1})}\right|\nonumber\\&\ge\mathcal{L}_{D}\left(\widetilde{\rho}(\tau_{2}),\widetilde{\rho}(\tau_{1})\right).
    \label{eq:monotonicity_ineq}
\end{align}
Equation~\eqref{eq:monotonicity_ineq} can be regarded as a quantum speed limit.

Using the fidelity, we can derive a quantum thermodynamic uncertainty relation. Let us consider an observable $\mathcal{C}$, which is a time-independent Hermitian operator.
We define the mean and the standard deviation with respect to $\ket{\tilde{\psi}}$:
\begin{align}
    \braket{\mathcal{C}}(t)&\equiv\mathrm{Tr}[\mathcal{C}\widetilde{\rho}(t)]=\braket{\widetilde{\psi}(t)|\mathcal{C}|\widetilde{\psi}(t)},\label{eq:def_mean}\\\Delta \mathcal{C}(t)&\equiv\sqrt{\braket{\mathcal{C}^{2}}(t)-\braket{\mathcal{C}}(t)^{2}}.
    \label{eq:stdev_def}
\end{align}
Then the quantum thermodynamic uncertainty relation gives a lower bound for the scaled standard deviation $\Delta \mathcal{C} / \braket{\mathcal{C}}$. 
One reason for considering such scaled variance is related to quantum clocks. 
Quantum clocks that repeatedly emit signals to their surroundings can be modeled using a continuous measurement formalism (cf. Section~\ref{sec:application}). This approach is necessary because these clocks interact with their environment through the process of signal emission. 
Quantum thermal machines
in which the system is coupled to thermal reservoirs are known to be used as clocks \cite{Erker:2017:QClockTUR,Mitchison:2019:QMachine}. By applying a continuous measurement formalism to such thermal machines, it becomes possible to model the system dynamics and the statistics of the emitted signals. When we denote the number of signals emitted as the operator $\mathcal{C}$, its scaled variance represents the accuracy of that clock. The limit of such accuracy can be quantified by the quantum thermodynamic uncertainty relation \cite{Hasegawa:2020:QTURPRL}.

A comment is in order on the term ``thermodynamic uncertainty relation''. 
In this paper, inequalities for the scaled variance, the variance by the square of the mean, are referred to as thermodynamic uncertainty relations. Typically, when referring to thermodynamic uncertainty relations, the thermodynamic cost is the entropy production. In recent years, inequalities bounded by dynamical activity have often been called kinetic uncertainty relations. However, considering that dynamical activity itself is also a thermodynamic quantity and that there are recently established relations bounded by quantities that are neither entropy production nor dynamical activity \cite{Hasegawa:2020:TUROQS,Prech:2024:ClockUR}, we collectively refer to inequalities for scaled variance as thermodynamic uncertainty relations.

Reference~\cite{Hasegawa:2021:QTURLEPRL} used the fidelity to derive a quantum thermodynamic uncertainty relation using the lower bound on the Hellinger distance \cite{Nishiyama:2020:HellingerBound}:
\begin{align}
    |\braket{\widetilde\psi(\tau_{2})|\widetilde\psi(\tau_{1})}|\le \left[\left(\frac{\langle \mathcal{C}\rangle(\tau_{1})-\langle \mathcal{C}\rangle(\tau_{2})}{\Delta\mathcal{C}(\tau_{1})+\Delta\mathcal{C}(\tau_{2})}\right)^2+1\right]^{-\frac{1}{2}},
    \label{eq:fidelity_TUR}
\end{align}
Using Eqs.~\eqref{eq:illustrative_bound} and \eqref{eq:fidelity_TUR}, we obtain
\begin{align}
    \Lambda(\tau_{1},\tau_{2})\le\left[\left(\frac{\langle\mathcal{C}\rangle(\tau_{1})-\langle\mathcal{C}\rangle(\tau_{2})}{\Delta\mathcal{C}(\tau_{1})+\Delta\mathcal{C}(\tau_{2})}\right)^{2}+1\right]^{-\frac{1}{2}}.
    \label{eq:TUR_Lambda}
\end{align}
Equations~\eqref{eq:monotonicity_ineq} and \eqref{eq:TUR_Lambda} demonstrate that, once a lower bound for the fidelity is obtained, we can derive quantum speed limits and thermodynamic uncertainty relations.

\section{Results\label{sec:results}}

As discussed in the previous section, the fidelity leads to quantum speed limits and thermodynamic uncertainty relations.
Hence, a distinct lower  bound  of the fidelity results in a unique quantum speed limit and thermodynamic uncertainty relation. 
The original Margolus-Levitin and Mandelstam-Tamm quantum speed limits concern the expectation and the standard deviation of the Hamiltonian, respectively.
We derive Margolus-Levitin and Mandelstam-Tamm quantum speed limits for non-Hermitian systems. 
The derived results are summarized in Table~\ref{tab:results_summary}.

\subsection{Margolus-Levitin type tradeoff relations\label{sec:ML_tradeoff}}

First, we consider the Margolus-Levitin type tradeoff relations,
which include terms comprising the average of operators [Eq.~\eqref{eq:ML_bound_def}]. 
Here, we assume that the non-Hermitian Hamiltonian $\mathcal{H}$ is time-independent. 
Furthermore, we assume that $H$ and $\Gamma$ commute, that is
\begin{align}
    [H, \Gamma]=0.
    \label{eq:commutative_condition}
\end{align}
Although this assumption appears to be restrictive, 
this commutation assumption is satisfied by
the dephasing model,
which descirbes a loss of coherence owing to interactions with the environment,
or quantum thermal machines, where the jump operators are limited to jumps between energy eigenstates. 
Still, when considering an initial state having coherence with respect to the energy eigenstates, the dynamics becomes different from the classical dynamics.
These models are detailed in Appendix~\ref{sec:models}. 

Because $H$ is Hermitian, $H - E_g\mathbb{I}_S$ admits the following eigendecomposition ($E_g$ in the minimum eigenvalue of $H$):
\begin{align}
    H - E_g \mathbb{I}_S = \sum_j \epsilon_j \ket{\epsilon_j}\bra{\epsilon_j},
    \label{eq:H_eigendecomp}
\end{align}
where $\epsilon_j$ and $\ket{\epsilon_j}$ are the eigenvalues and eigenvectors, respectively, of $H - E_g \mathbb{I}_S$ (therefore, $\ket{\epsilon_j}$ is the eigenvector of $H$ as well). 
By definition, $\{\epsilon_j\}$ are nonnegative. 
 Since $[H, \Gamma] = 0$, we can choose $\{\ket{\epsilon_j}\}$ as eigenvectors of $\Gamma$, and let $\gamma_j$ be an eigenvalue corresponding to $\ket{\epsilon_j}$.  
We assume that $\gamma_j\geq 0$ for all $j$ (that is, $\Gamma$ is semipositive definite), which is satisfied for the Feshbach or quantum trajectory approach \cite{Ashida:2020:NonHermiteReview}.  
From now on, when
calculating expected values, we will omit the $\mathbb{I}_S$ if it is obvious. 

The original Margolus-Levitin quantum speed limit \cite{Margolus:1998:QSL} used the trigonometric inequality to derive the relation. 
Later, Refs.~\cite{Jones:2010:QSL,Zwierz:2012:CommentOnGQSL} showed that the Margolus-Levitin-type speed limit can be obtained via geometric consideration, where the fidelity between two quantum states play a central role. 
We apply the procedure in Ref.~\cite{Zwierz:2012:CommentOnGQSL} to the system having a non-Hermitian Hamiltonian, which leads to Eq.~\eqref{eq:weaker_ML_bound_def} for an isolated quantum dynamics. 

For $\lambda\in \mathbb{R}$, we define the state vector as
\begin{align}
    \label{eq:def_psi_t_lambda}
    \ket{\psi_\lambda(t)}\equiv (e^{-i(\mathcal{H}-\lambda)t}\otimes\mathbb{I}_{A})\ket{\psi(0)}=e^{i\lambda t}\ket{\psi(t)}.
\end{align}

As mentioned in Section~\ref{sec:fidelity_tradeoff}, we are interested in the  the fidelity $\left|\braket{\widetilde{\psi}(0)|\widetilde{\psi}(\tau)}\right|$.
The fidelity is bounded from below by
\begin{align}
    \left|\braket{\widetilde{\psi}(0)|\widetilde{\psi}(\tau)}\right|&=\frac{|\braket{\psi_{E_{g}}(0)|\psi_{E_{g}}(\tau)}|}{\left\Vert \ket{\psi(\tau)}\right\Vert }\nonumber\\&\ge\Lambda_{\mathrm{ML}}(0,\tau),
    \label{eq:fidelity_main}
\end{align}
where $\Lambda_\mathrm{ML}(0,\tau)$ is the lower bound given in Eq.~\eqref{eq:illustrative_bound} for the Margolus-Levitin type tradeoff relations:
\begin{align}
    \Lambda_{\mathrm{ML}}(0,\tau)\equiv\frac{e^{-\braket{\Gamma}(0)\tau}-\tau\left(\braket{H}(0)-E_{g}\right)}{\left\Vert \ket{\psi(\tau)}\right\Vert }.
    \label{eq:Lambda_ML_def}
\end{align}
For details of the derivation of Eqs.~\eqref{eq:fidelity_main} and \eqref{eq:Lambda_ML_def}, please see Appendix~\ref{sec:fidelity_deriv_upperbound}. 
Using the montonicity [Eq.~\eqref{eq:Bures_monotonicity}], from Eq.~\eqref{eq:fidelity_main}, we obtain a quantum speed limit:
\begin{align}
    &1-\sqrt{\mathrm{Fid}(\widetilde{\rho}(0)), \widetilde{\rho}(\tau))}=2\sin^{2}\left(\frac{\mathcal{L}_{D}(\widetilde{\rho}(0),\widetilde{\rho}(\tau))}{2}\right)\nonumber\\&\le1+\frac{\tau\left(\braket{H}(0)-E_{g}\right)-e^{-\braket{\Gamma}(0)\tau}}{\|\ket{\psi(\tau)}\|}.
    \label{eq:QSL_ML}
\end{align}
Because $\sin(x)^2$ is an increasing function for $0\le x \le \pi/2$,
it is possible to make more significant changes to the state provided the system with $H$ and $\Gamma$ having larger expectation values. 
Since $0 \le \mathrm{Fid}(\widetilde{\rho}(0)), \widetilde{\rho}(\tau))\le 1$,
Eq.~\eqref{eq:QSL_ML} is nontrivial when $e^{-\braket{\Gamma}(0)\tau}-\tau\left(\braket{H}(0)-E_{g}\right) > 0$. 
Assuming this condition, since $\|\ket{\psi(\tau)}\|=\braket{\psi(0)|e^{-2\Gamma \tau}|\psi(0)}\le 1$ from $[H, \Gamma]=0$, we obtain the simpler relation:
\begin{align}
    \label{eq:ML_like_bound_simple_main}
     &2\sin^{2}\left(\frac{\mathcal{L}_{D}(\widetilde{\rho}(0),\widetilde{\rho}(\tau))}{2}\right)\nonumber\\&\le\tau\left(\braket{H}(0)-E_{g}\right)+1-e^{-\braket{\Gamma}(0)\tau}\nonumber\\&\le\tau\left(\braket{H}(0)-E_{g}+\braket{\Gamma}(0)\right).
\end{align}
When $\Gamma = 0$ (i.e., an isolated dynamics)  and density matrices are fully distinguishable (i.e., $\mathrm{Fid}(\widetilde{\rho}(0), \widetilde{\rho}(\tau))=0$) ,  Eq.~\eqref{eq:ML_like_bound_simple_main} becomes identical to Eq.~\eqref{eq:weaker_ML_bound_def}.

Using $\Lambda_\mathrm{ML}(0,\tau)$ shown in Eq.~\eqref{eq:Lambda_ML_def}, we can derive a thermodynamic uncertainty relation from Eq.~\eqref{eq:TUR_Lambda}. 
Again, assuming $e^{-\braket{\Gamma}(0)\tau}-\tau\left(\braket{H}(0)-E_{g}\right) > 0$,
we obtain
\begin{align}
    &\frac{1}{\left[e^{-\tau\braket{\Gamma}(0)}-\tau\left(\braket{H}(0)-E_{g}\right)\right]^{2}}-1\nonumber\\&\geq\frac{\|\ket{\psi(\tau)}\|^2}{\left[e^{-\tau\braket{\Gamma}(0)}-\tau\left(\braket{H}(0)-E_{g}\right)\right]^{2}}-1\nonumber\\&\geq\left(\frac{\langle\mathcal{C}\rangle(\tau)-\langle\mathcal{C}\rangle(0)}{\Delta\mathcal{C}(\tau)+\Delta\mathcal{C}(0)}\right)^{2}.
    \label{eq:ML_type_UR_main}
\end{align}
In Eq.~\eqref{eq:ML_type_UR_main}, the term $\tau (\braket{H}(0) - E_g)$ is always positive for a quantum dynamics. 
As in the speed limit of Eq.~\eqref{eq:QSL_ML}, 
the accuracy of the observable increases when the system with $H$ and $\Gamma$ exhibits higher expectation values.

\subsection{Mandelstam-Tamm type tradeoff relations\label{sec:MT_tradeoff}}

Next, we consider the Mandelstam-Tamm-type tradeoff relations,
 that is, where the cost part includes the standard deviation of operators [Eq.~\eqref{eq:MT_bound_def}].

The original derivation of the Mandelstam-Tamm quantum speed limit was based on the Heisenberg-Robertson uncertainty relation.
Here, we follow the procedures developed in Refs.~\cite{vaidman1992minimum, PhysRevA.104.052620}.
Reference~\cite{vaidman1992minimum} proved the Mandelstam-Tamm quantum speed limit considering the time-derivative of the fidelity in an isolated quantum system (i.e., Hermitian Hamiltonians).
Reference~\cite{PhysRevA.104.052620} extended the approach of  Ref.~\cite{vaidman1992minimum} to non-Hermitian Hamiltonians. 
For integrity of the paper, we review the results of Ref.~\cite{PhysRevA.104.052620}. 

The fidelity is bounded from below by 
\begin{align}
    |\braket{\widetilde{\psi}(\tau_{1})|\widetilde{\psi}(\tau_{2})}|\geq \cos\left[\int_{\tau_{1}}^{\tau_{2}}dt\Delta\mathcal{H}(t)\right]\equiv \Lambda_{\mathrm{MT}}(\tau_{1},\tau_{2}),
    \label{eq:Lambda_MT_def}
\end{align}
where $\Lambda_{\mathrm{MT}}$ is the lower bound given in in Eq.~\eqref{eq:illustrative_bound} for the Mandelstam-Tamm type tradeoff relations.
In Eq.~\eqref{eq:Lambda_MT_def}, we assumed $0 \le \int_{\tau_{1}}^{\tau_{2}}dt\Delta\mathcal{H}(t) \le \pi / 2$. 
For details of the derivation of Eq.~\eqref{eq:Lambda_MT_def}, please see Appendix~\ref{sec:MT_fidelity_deriv}. 
Applying the monotonicity relation to  Eq.~\eqref{eq:Lambda_MT_def}, we obtain the quantum speed limit:
\begin{align}
    \int_{\tau_{1}}^{\tau_{2}}dt\Delta\mathcal{H}(t)\geq\mathcal{L}_{D}\left(\widetilde{\rho}(\tau_{1}),\widetilde{\rho}(\tau_{2})\right).
    \label{eq:MT_QSL}
\end{align}
Quantum speed limits are usually described as inequalities that provide lower bounds for the elapsed time.
By expressing Eq.~\eqref{eq:MT_QSL} in this way, 
Eq.~\eqref{eq:MT_QSL} can be equivalently given by
\begin{align}
    \tau\geq\frac{\mathcal{L}_{D}\left(\widetilde{\rho}(0),\widetilde{\rho}(\tau)\right)}{\overline{\Delta\mathcal{H}}},
    \label{eq:QSL_as_tau_lowerbound}
\end{align}
where $\overline{\Delta\mathcal{H}}\equiv \frac{1}{\tau}\int_{0}^{\tau}dt\Delta\mathcal{H}(t)$ is the temporal average of the  standard deviation  of $\mathcal{H}$. 
Equation~\eqref{eq:QSL_as_tau_lowerbound} generalizes the expression of Eq.~\eqref{eq:MT_bound_def} for non-Hermitian dynamics.

Similarly, using Eqs.~\eqref{eq:TUR_Lambda} and \eqref{eq:Lambda_MT_def}, the corresponding thermodynamic uncertainty relation becomes
\begin{align}
    \label{eq:MT_Type_UR}
    \left[\tan\left(\int_{\tau_1}^{\tau_2} dt \Delta \mathcal{H}(t) \right)\right]^2\geq \left(\frac{\langle \mathcal{C}\rangle(\tau_2)-\langle \mathcal{C}\rangle(\tau_1)}{\Delta\mathcal{C}(\tau_2)+\Delta\mathcal{C}(\tau_1)}\right)^2.
\end{align}
Moreover, we can derive a thermodynamic uncertainty relation for a short time limit. 
Let $\delta > 0$ be an infinitesimally small value. 
Taking $\tau_1=\tau-\delta$ and $\tau_2=\tau$, we obtain
\begin{align}
    \label{eq:MT_Type_UR2}
    \Delta \mathcal{H}(\tau)^2 \delta^2 \geq \left(\frac{d_t\langle \mathcal{C}\rangle(\tau)\delta}{2\Delta\mathcal{C}(\tau)}\right)^2 + O(\delta^3),
\end{align}
where $d_t$ denotes the differential with respect to $t$.
Hence, we obtain
\begin{align}
    \label{eq:robertson_like}
    \Delta \mathcal{C}(\tau)\Delta\mathcal{H}(\tau)\geq \frac{1}{2}\left|d_t \braket{\mathcal{C}}(\tau)\right|,
\end{align}
which corresponds to the energy-time uncertainty relation for the non-Hermitian Hamiltonian.

\section{Example: Continuous measurement\label{sec:application}}

\begin{figure}
    \includegraphics[width=1.0\linewidth]{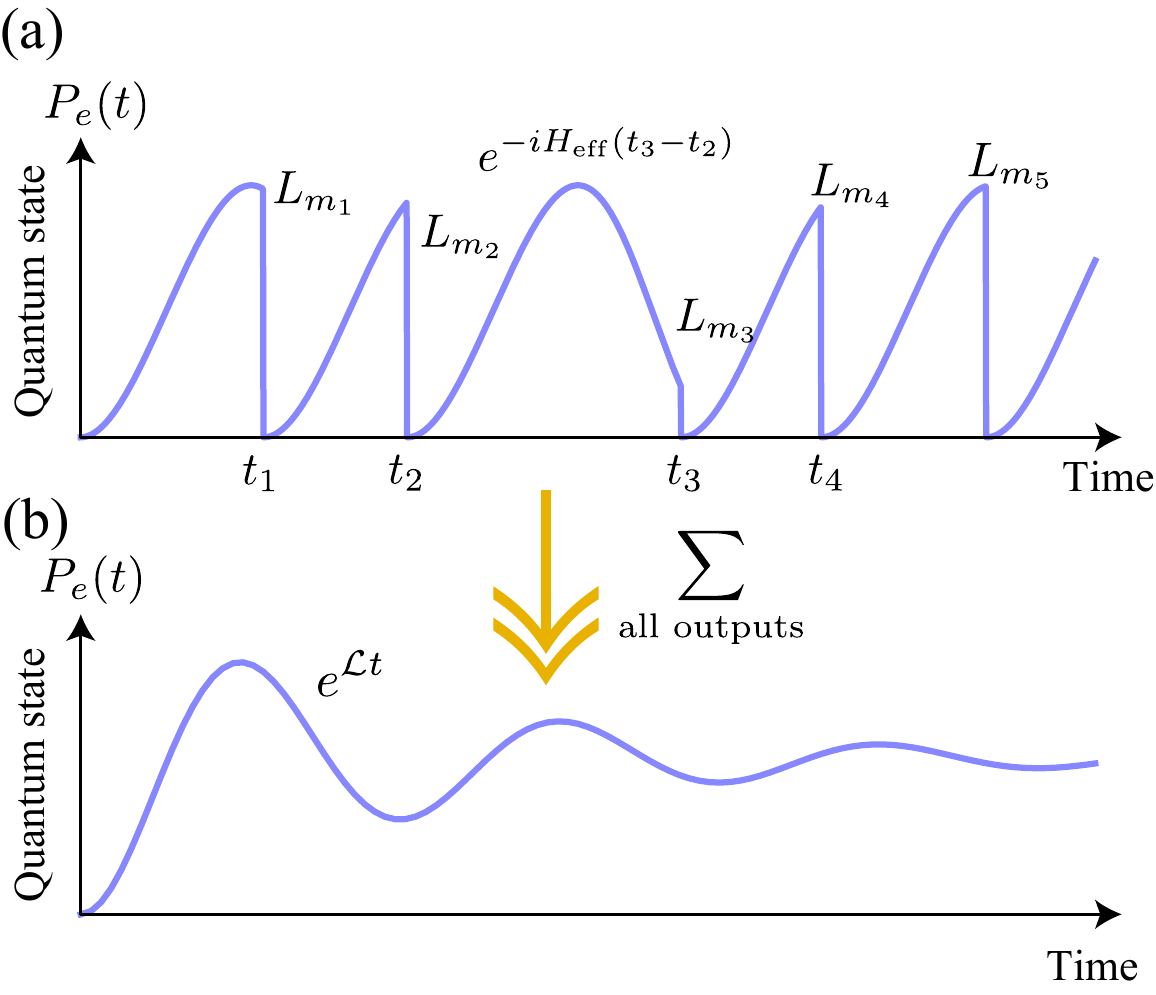}
    \caption{
    (a) Example of quantum trajectory, where the horizontal and vertical axes denote time and $P_e(t) = \mathrm{Tr}_S[\rho_S(t)\ket{\mathfrak{e}}\bra{\mathfrak{e}}]$ ($\ket{\mathfrak{e}}$ is the excited state). 
    Here, $t_k$ denotes the time of the $k$th jump event. 
    The trajectory comprises two type of dynamics. The first is given by an effective Hamiltonian $\exp(-iH_\mathrm{eff}(t_{k}-t_{k-1}))$, which induces a consistent, smooth transition in the state. The second factor is the jump operator term  $L_m$. Contrary to the first, this factor causes a sudden, abrupt jump in the state.
    (b) Trajectory when averaging over all possible measurement outputs. The dynamics obeys the Lindblad equation [Eq.~\eqref{eq:Lindblad_eq_def}]. The time evolution is $e^{\mathcal{L}t}$ which is different from $e^{-iH_\mathrm{eff}t}$.  
}
    \label{fig:quantum_trajectory}
\end{figure}

In the preceding sections, we have derived
quantum speed limits and thermodynamic uncertainty relations for a quantum system having a non-Hermitian Hamiltonian.
In this section, for a specific example, we consider a quantum system that undergoes continuous measurement,
which includes dynamics induced by non-Hermtian Hamiltonians. 
Our focus is on a quantum Markov process, described by the Lindblad equation. Let $\mathcal{L}$ be the Lindblad superoperator. The Lindblad equation can then be expressed as
\begin{align}
     \dot{\rho}_S=\mathcal{L} \rho_S\equiv -i[H_S, \rho_S]+\sum_{m=1}^{N_C} \mathcal{D}[L_m] \rho_S.
     \label{eq:Lindblad_eq_def}
\end{align}
Here, $H_S$ denotes the system Hamiltonian, $L_m$ denotes the $m$th jump operator where $m$ ranges from $1$ to $N_C$ (the number of channels), and $\mathcal{D}[L]\rho_S = L\rho_SL^\dagger - \frac{1}{2}\{L^\dagger L, \rho_S\}$ is the dissipator describing the interaction between the system and the environment.
The Lindblad equation can simulate both classical Markov processes and closed quantum evolution. Setting $H_S$ to 0 yields the classical Markov process, while setting $L_m$ to 0 for all $m$ corresponds to an isolated quantum dynamics. 
Equation~\eqref{eq:Lindblad_eq_def} can be represented by
\begin{align}
    \dot{\rho}_S=-i\left(H_{\mathrm{eff}} \rho_S-\rho_S H_{\mathrm{eff}}^{\dagger}\right)+\sum_{m=1}^{N_C} L_m \rho_S L_m^{\dagger},
    \label{eq:Lindblad_eq_def2}
\end{align}
where $H_\text{eff} \equiv H_S - \frac{i}{2}\sum_m L_m^\dagger L_m$ is the effective Hamiltonian which is a non-Hermitian operator. 
According to Eq.~\eqref{eq:Lindblad_eq_def2}, it can be seen that the state change in the Lindblad equation comprises the smooth state change by the effective Hamiltonian $H_\mathrm{eff}$ and the discontinuous change by the jump operators $L_m$.
The Lindblad equation describes the dynamics in situations where the environment is not being monitored.
When we measure the environment, depending on the outcome of the measurement, the state of the system exhibits stochastic dynamics, which is called a quantum trajectory.
A quantum trajectory comprises the smooth dynamics between two jump events that is governed by the non-Hermitian Hamiltonian (Fig.~\ref{fig:quantum_trajectory}(a)). 
When averaging over all possible quantum trajectories, the dynamics of the resulting density operator obeys the original Lindblad equation (Fig.~\ref{fig:quantum_trajectory}(b)).

The Lindblad equation can also be represented using the Kraus representation:
\begin{align}
    \rho_S(t+dt) = \sum_{m=0}^{N_C} V_m\rho_S(t)V_m^\dagger.
    \label{eq:rhoS_Kraus_repr}
\end{align}
where $V_m$ is the Kraus operator defined as $V_0 = \mathbb{I}_S - idt H_\text{eff}$ and $V_m = \sqrt{dt}L_m$ for $1 \leq m \leq N_C$.
The Kraus operator should satisfy the completeness relation $\sum_{m=0}^{N_C} V_m^\dagger V_m = \mathbb{I}_S$.
Applying the Kraus representation repeatedly, we can express $\rho_S(\tau)$ 
by
\begin{align}
    \rho_{S}(\tau)=\sum_{\boldsymbol{m}}V_{m_{K-1}}\cdots V_{m_{0}}\rho_{S}(0)V_{m_{0}}^{\dagger}\cdots V_{m_{K-1}}^{\dagger},
    \label{eq:rhoS_Kraus_repr_tau}
\end{align}
where $K$ is sufficiently large natural number and $dt = \tau / K$. 
From the Kraus representation of Eq.~\eqref{eq:rhoS_Kraus_repr_tau}, we can introduce the following matrix product state (MPS):
\begin{align}
    \ket{\Phi(\tau)}&=\sum_{\boldsymbol{m}}V_{m_{K-1}}\ldots V_{m_{0}}\ket{\psi_{S}(0)}\otimes\ket{\boldsymbol{m}}.
    \label{eq:MPS_def}
\end{align}
Further, taking $dt \to 0$ limit in Eq.~\eqref{eq:MPS_def} keeping $\tau$ constant, we can express $\ket{\Phi(\tau)}$ 
using the continuous MPS:
\begin{align}
\ket{\Phi(\tau)}=\mathcal{U}(\tau)\ket{\psi_{S}(0)}\otimes\ket{\mathrm{vac}},
\label{eq:cMPS_def}
\end{align}
where the operator $\mathcal{U}(\tau)$ is defined as follows:
\begin{align}
\mathcal{U}(\tau)=\mathbb{T}e^{-i\int_{0}^{\tau}dt\,(H_{\mathrm{eff }}\otimes\mathbb{I}_{F}+\sum_{m}iL_{m}\otimes\phi_{m}^{\dagger}(t))}.
\label{eq:mathcalV_def}
\end{align}
In the equations above, $\mathbb{T}$ represents the time-ordering operator, $\mathbb{I}_F$ is the field identity operator, and $\phi_m(s)$ is the field operator that fulfills the canonical communication relation. The vacuum state is denoted as $|\mathrm{vac}\rangle$. 
The evaluation of fidelity between two states at different times,  $\braket{\Phi(t_2)|\Phi(t_1)}$,  is difficult with the continuous MPS provided by Eq.~\eqref{eq:cMPS_def}. This is because assessing the fidelity between two continuous MPS at different times is not feasible. Considering these constraints, we use the scaled continuous MPS \cite{Hasegawa:2023:BulkBoundaryBoundNC}:
\begin{align}
\ket{\Psi(t)}\equiv\mathcal{V}(\theta)\ket{\psi_{S}(0)}\otimes\ket{\mathrm{vac}},
\label{eq:scaled_cMPS}
\end{align}
where $\theta \equiv t / \tau$ is the scaled time and $\mathcal{V}(\theta)$ is defined by
\begin{align}
\mathcal{V}(\theta)\equiv\mathbb{T}e^{\int_{0}^{\tau}ds\,(-i\theta H_{\mathrm{eff}}\otimes\mathbb{I}_{\mathrm{F}}+\sqrt{\theta}\sum_{m}L_{m}\otimes\phi_{m}^{\dagger}(s))}.
\label{eq:mathcalV_theta_def}
\end{align}
Let us consider the fidelity between two scaled cMPS states, $\braket{\Psi(0)|\Psi(t)}$.
Recall that 
\begin{align}
    \braket{\Psi(0)|\Psi(t)}&=\mathrm{Tr}_{SAF}[\ket{\Psi(t)}\bra{\Psi(0)}]\nonumber\\
    &=\mathrm{Tr}_{S}\left[\mathrm{Tr}_{AF}[\ket{\Psi(t)}\bra{\Psi(0)}]\right]\nonumber\\
    &=\mathrm{Tr}_{S}\left[\varrho(t)\right],
    \label{eq:Tr_SAF}
\end{align}
where $\mathrm{Tr}_{F}$ is the trace with respect to the field and $\varrho(t) \equiv \mathrm{Tr}_{AF}[\ket{\Psi(t)}\bra{\Psi(0)}]$. 
Because $\phi_m(s)$ annihilates the vacuum and $\ket{\Psi(0)}=\ket{\psi_S(0)}\otimes\ket{\mathrm{vac}}$,
the time evolution of $\varrho(t)$ from $t$ to $t+dt$ is
\begin{align}
    \varrho(t+dt)&=\mathrm{Tr}_{AF}\left[\ket{\Psi(t+dt)}\bra{\Psi(0)}\right]\nonumber\\&=\left(\mathbb{I}_{S}-idtH_{\mathrm{eff}}\right)\mathrm{Tr}_{AF}\left[\ket{\Psi(t)}\bra{\Psi(0)}\right]\nonumber\\&=\left(\mathbb{I}_{S}-idtH_{\mathrm{eff}}\right)\varrho(t),
    \label{eq:varrho_onestep}
\end{align}
which implies the following differential equation:
\begin{align}
    \dot{\varrho}=-iH_{\mathrm{eff}}\varrho.
    \label{eq:varrho_ode}
\end{align}

Until now, we assumed that $H_S$ and $L_m$ do not depend on time $t$. 
It is straightforward to generalize the discussion to consider time-dependent $H_S$ and $L_m$. Then, the fidelity becomes
\begin{align}
    \label{eq:Psi_psi_main}
    \braket{\Psi(0)|\Psi(t)}&=\mathrm{Tr}_S[\mathbb{T}e^{-i\int_0^t ds H_{\mathrm{eff}}(s)}\rho_S(0)]\nonumber\\
    &=\braket{\psi_{S}0)|\mathbb{T}e^{-i\int_0^t ds H_{\mathrm{eff}}(s)}|\psi_{S}(0)}.
\end{align}

In a similar way as in Eq.~\eqref{eq:def_psi_t_lambda}, we define
\begin{align}
    \label{eq:def_psi_hat_open_main}
    \ket{\hat\psi_\lambda(t)}\equiv (\mathbb{T}e^{-i\int_0^t ds \left(H_{\mathrm{eff}}(s)-\lambda\right)}\otimes\mathbb{I}_A) \ket{\psi_{S}(0)}.
\end{align}
By combining this relation with Eq.~\eqref{eq:Psi_psi_main}, we obtain
\begin{align}
    |\braket{\Psi(0)|\Psi(t)}|=|\braket{\hat\psi_\lambda(0)|\hat\psi_\lambda(t)}|.
     \label{eq:Psi_psi_absolute_main}
\end{align}
By applying the following replacement
\begin{align}
    H&=H_{S},\label{eq:H_HS_replace}\\\Gamma&=\frac{1}{2}\sum_{m}L_{m}^{\dagger}L_{m}.\label{eq:Gamma_replace}
\end{align}
we can apply the method discussed above.

\subsection{Margolus-Levitin type tradeoff relations\label{subsec:app_ML}}

Let us first consider Margolus-Levitin type tradeoff relations in the continuous measurement formalism. 
 Suppose  $H_S$ and $\{L_m\}$ are time independent.
Since we required the commutative condition given by Eq.~\eqref{eq:commutative_condition} for the Margolus-Levitin quantum speed limit, 
we assume that $H_S$ and $\sum_m L_m^\dagger L_m$ commute, i.e., $[H_S, \sum_m L_m^\dagger L_m]=0$. 
For instance, this condition is satisfied by the dephasing model  (Appendix~\ref{sec:dephasing_model}) or  autonomous refrigerator  (Appendix~\ref{sec:quantum_heat_engine}).
Let $E_{S,g}\in\mathbb{R}$ be the minimum eigenvalue of $H_S$. By combining Eq.~\eqref{eq:fidelity_main} and Eq.~\eqref{eq:Lambda_ML_def} with Eq.~\eqref{eq:Psi_psi_absolute_main} for $\lambda=E_{S,g}$, we obtain 
\begin{align}
    |\braket{\Psi(0)|\Psi(\tau)}|\geq e^{-\frac{1}{2}\mathfrak{a}(0)\tau}-\tau\left(\braket{H_S}(0) -E_{S,g}\right),
    \label{eq:fidelity_open_main}
\end{align}
where $\mathfrak{a}(0)\equiv \mathrm{Tr}_{S}\left[\sum_m L_m^\dagger L_m\rho_S(0)\right]$, $\braket{\bullet}(t)\equiv \braket{\Psi(t)|\bullet |\Psi(t)}$ and we use 
$\braket{\Psi(0)|\bullet |\Psi(0)}=\braket{\psi_{S}(0)|\bullet |\psi_{S}(0)}$.
Hence, the lower bound for the fidelity corresponding the Margolus-Levitin type case is given by 
\begin{align}
    \hat{\Lambda}_{\mathrm{ML}}(0,\tau)=e^{-\frac{1}{2}\mathfrak{a}(0)\tau}-\tau\left(\braket{H_S}(0) -E_{S,g}\right).
    \label{eq:Lambda_ML_open_def}
\end{align}
where the hat is added to emphasize that $\Lambda_\mathrm{ML}$ is defined for the continuous measurement formalism. 
Hereafter, when the hat is added to an operator, it means that it is an operator for the continuous measurement.
Repeating the calculation in Section~\ref{sec:ML_tradeoff}, we obtain the quantum speed limit:
\begin{align}
    &1 +\tau \left(\braket{H_S}(0) -E_{S,g}\right) - e^{-\frac{1}{2}\mathfrak{a}(0)\tau}\nonumber\\
    &\geq 2\sin^2\left(\frac{\mathcal{L}_D(\rho_S(0), \rho_S(\tau))}{2}\right),
     \label{eq:ML_bound_open_main}
\end{align}
and the thermodynamic uncertainty relation:
\begin{align}
&\frac{1}{\left[e^{-\frac{1}{2}\mathfrak{a}(0)\tau}-\tau\left(\braket{H_{S}}(0)-E_{S,g}\right)\right]^{2}}-1\nonumber\\&\geq\left(\frac{\langle\hat{\mathcal{C}}\rangle(\tau)-\langle\hat{\mathcal{C}}\rangle(0)}{\Delta\hat{\mathcal{C}}(\tau)+\Delta\hat{\mathcal{C}}(0)}\right)^{2},
     \label{eq:ML_type_UR_open_main}
\end{align}
where we assume $e^{-\mathfrak{a}(0)\tau/2}-\tau\left(\braket{H_S}(0)-E_{S,g}\right) > 0$.
Notably, Eq.~\eqref{eq:ML_type_UR_open_main} can be represented by the dynamical activity and the expectation value of the Hamiltonian. 
The thermodynamic uncertainty relation of Eq.~\eqref{eq:ML_type_UR_open_main} shows that, in the presence of coherent dynamics, the left-hand side of Eq.~\eqref{eq:ML_type_UR_open_main} becomes larger. 

In the classical case,
$H_S=0$ and thus
the commutation condition $[H_S, \sum_m L_m^\dagger L_m]=0$ is automatically satisfied, so the Margolus-Levitin type tradeoff relations hold. The classical case is discussed in Appendix~\ref{sec:classical_case}.

\subsection{Mandelstam-Tamm type tradeoff relations}

Next, we consider the continuous measurement for
the Mandelstam-Tamm type tradeoff relations. 
We define
\begin{align}
    \ket{\hat\psi(t)}\equiv \ket{\hat\psi_{\lambda = 0}(t)}=(\mathbb{T}e^{-i\int_0^t ds H_{\mathrm{eff}}(s)}\otimes \mathbb{I}_A)\ket{\psi_{S}(0)},
    \label{eq:ket_hat_psi_def}
\end{align}
where $\ket{\psi_\lambda(t)}$ is defined in Eq.~\eqref{eq:def_psi_t_lambda}. 
We proceed by defining the \textit{pseudo density operator} as well as its associated variance:
\begin{align}
    \rho_{\mathrm{P}}(t)&\equiv\frac{1}{Z(t)}\mathbb{T}e^{-i\int_{0}^{t}dsH_{\mathrm{eff}}(s)}\rho_{S}(0)\overline{\mathbb{T}}e^{i\int_{0}^{t}dsH_{\mathrm{eff}}^{\dagger}(s)}\nonumber\\&=\mathrm{Tr}_{A}[\ket{\widetilde{\hat{\psi}}(t)}\bra{\widetilde{\hat{\psi}}(t)}],
     \label{eq:def_trace_main}
\end{align}
where 
$\ket{\widetilde{\hat{\psi}}(t)}\equiv\ket{\hat{\psi}(t)}/\|\ket{\hat{\psi}(t)}\|$ is the normalized vector and 
\begin{align}
    Z(t)&\equiv\mathrm{Tr}_{S}\left[\mathbb{T}e^{-i\int_{0}^{t}dsH_{\mathrm{eff}}(s)}\rho_{S}(0)\overline{\mathbb{T}}e^{i\int_{0}^{t}dsH_{\mathrm{eff}}^{\dagger}(s)}\right]\nonumber\\&=\braket{\hat{\psi}(t)|\hat{\psi}(t)}.
    \label{eq:Zt_def}
\end{align}
We define the standard deviation of $H_\mathrm{eff}$ for the pseudo density operator as follows:
\begin{align}
    &\Delta_{\mathrm{P}}H_{\mathrm{eff}}(t)\nonumber\\&\equiv\sqrt{\mathrm{Tr}_{S}[H_{\mathrm{eff}}^{\dagger}(t)H_{\mathrm{eff}}(t)\rho_{\mathrm{P}}(t)]-\left|\mathrm{Tr}_{S}[H_{\mathrm{eff}}(t)\rho_{\mathrm{P}}(t)]\right|^{2}}\nonumber\\&=\sqrt{\braket{\widetilde{\hat{\psi}}(t)|H_{\mathrm{eff}}^{\dagger}(t)H_{\mathrm{eff}}(t)|\widetilde{\hat{\psi}}(t)}-\left|\braket{\widetilde{\hat{\psi}}(t)|H_{\mathrm{eff}}(t)|\widetilde{\hat{\psi}}(t)}\right|^{2}}.
    \label{eq:def_normalized_density_main}
\end{align}
This represents the standard deviation of the effective Hamiltonian $H_\mathrm{eff}$ under the condition that no jumps are present.
From Eq.\eqref{eq:def_trace_main}, note that the pseudo density matrix have unit trace. 
 By combining Eq.~\eqref{eq:Lambda_MT_def} 
with Eq.~\eqref{eq:Psi_psi_absolute_main} for $\lambda=0$, we have
\begin{align}
    &|\braket{\Psi(0)|\Psi(\tau)}|=\sqrt{Z(\tau)}|\braket{\widetilde{\hat\psi}(0)|\widetilde{\hat\psi}(\tau)}|\nonumber\\
    &\geq \sqrt{Z(\tau)}\cos\left(\int_0^\tau dt \Delta_\mathrm{P} H_{\mathrm{eff}}(t)\right).
    \label{eq:abs_inner_product}
\end{align}
Hence, the lower bound for the fidelity corresponding to the Mandelstam-Tamm type case is given by  
\begin{align}
    \hat{\Lambda}_{\mathrm{MT}}(0,\tau)=\sqrt{Z(\tau)}\cos\left(\int_0^\tau dt \Delta_\mathrm{P} H_{\mathrm{eff}}(t)\right).
    \label{eq:Lambda_hat_MT}
\end{align}
Applying the monotonicity relation, we obtain a quantum speed limit:
\begin{align}
    &\int_0^\tau dt \Delta_\mathrm{P} H_{\mathrm{eff}}(t) 
    \geq \arccos\left(\sqrt{\frac{\mathrm{Fid}\left(\rho_S(0), \rho_S(\tau)\right)}{Z(\tau)}}\right).
    \label{eq:LM_type_fidelity_open_main}
\end{align}
In the case without a jump operator, it reduces to the well-known quantum speed limit.
Repeating the calculation in Section~\ref{sec:MT_tradeoff}, we obtain an uncertainty relation:
\begin{align}
    \frac{1}{Z(\tau)\left[\cos\left(\int_0^\tau dt \Delta_\mathrm{P} H_{\mathrm{eff}}(t) \right)\right]^2}-1\geq \left(\frac{\langle \hat{\mathcal{C}}\rangle(\tau)-\langle \hat{\mathcal{C}}\rangle(0)}{\Delta\hat{\mathcal{C}}(\tau)+\Delta\hat{\mathcal{C}}(0)}\right)^2.
    \label{eq:MT_Type_UR_open_main}
\end{align}

\section{Concluding remarks}

This study presents a unified framework for deriving trade-off relations, including quantum speed limits and thermodynamic uncertainty relations, for systems governed by non-Hermitian Hamiltonians. 
We extended the Margolus-Levitin and Mandelstam-Tamm quantum speed limits, originally derived for isolated quantum dynamics, to systems with non-Hermitian Hamiltonians. 
Furthermore, we derived bounds on the ratio of the standard deviation to the mean of an observable, which have the same form as the thermodynamic uncertainty relation.
By establishing a link between the fidelity and the derivation of the quantum speed limit and thermodynamic uncertainty relations, we showed that their framework extends the results of previous studies and provides a more comprehensive understanding of these fundamental tradeoff relations.
To demonstrate the applicability of the derived bounds, we apply them to the continuous measurement formalism in open quantum dynamics, where the dynamics is described by discontinuous jumps and smooth evolution induced by the non-Hermitian Hamiltonian. 
This work provides a unified perspective on quantum speed limits and thermodynamic uncertainty relations from the viewpoint of non-Hermitian Hamiltonians, paving the way for further advances in this research area.

\begin{acknowledgements}
This work was supported by Japan Society for the Promotion of Science KAKENHI Grant No. JP23K24915.
\end{acknowledgements}

\appendix

\begin{widetext}
\section{Derivation of Eq.~\eqref{eq:fidelity_main}\label{sec:fidelity_deriv_upperbound}}
Following Ref.~\cite{Zwierz:2012:CommentOnGQSL}, 
we calculate the time derivative of the fidelity $\left|\braket{\psi_{\lambda}(0)|\psi_{\lambda}(t)}\right|$.
\begin{align}
    \label{eq:derivative_fidelity_appendix}
   \left|\frac{d}{dt}|\braket{\psi_{\lambda}(0)|\psi_{\lambda}(t)}|\right|&=\frac{\left|\braket{\psi_{\lambda}(t)|\psi_{\lambda}(0)}\braket{\psi_{\lambda}(0)|d_{t}\psi_{\lambda}(t)}+\braket{d_{t}\psi_{\lambda}(t)|\psi_{\lambda}(0)}\braket{\psi_{\lambda}(0)|\psi_{\lambda}(t)}\right|}{2\left|\braket{\psi_{\lambda}(0)|\psi_{\lambda}(t)}\right|}\nonumber\\&=\frac{\left|-i\braket{\psi_{\lambda}(t)|\psi_{\lambda}(0)}\braket{\psi_{\lambda}(0)|(\mathcal{H}-\lambda)|\psi_{\lambda}(t)}+i\braket{\psi_{\lambda}(t)|(\mathcal{H}^{\dagger}-\lambda)|\psi_{\lambda}(0)}\braket{\psi_{\lambda}(0)|\psi_{\lambda}(t)}\right|}{2\left|\braket{\psi_{\lambda}(0)|\psi_{\lambda}(t)}\right|}\nonumber\\&=\frac{\left|\mathrm{Im}\left(\braket{\psi_{\lambda}(t)|\psi_{\lambda}(0)}\braket{\psi_{\lambda}(0)|(\mathcal{H}-\lambda)|\psi_{\lambda}(t)}\right)\right|}{\left|\braket{\psi_{\lambda}(0)|\psi_{\lambda}(t)}\right|}\nonumber\\&\le\frac{\left|\left(\braket{\psi_{\lambda}(t)|\psi_{\lambda}(0)}\braket{\psi_{\lambda}(0)|(\mathcal{H}-\lambda)|\psi_{\lambda}(t)}\right)\right|}{\left|\braket{\psi_{\lambda}(0)|\psi_{\lambda}(t)}\right|}\nonumber\\&=\left|\braket{\psi_{\lambda}(0)|(\mathcal{H}-\lambda)|\psi_{\lambda}(t)}\right|,
\end{align}
where $\ket{d_t\psi_\lambda(t)}$ denotes $d/dt\ket{\psi_\lambda(t)}$.
Substituting Eq.~\eqref{eq:nonHermitianHamiltonian_decompose} into Eq.~\eqref{eq:derivative_fidelity_appendix} with $\lambda = E_g$, we obtain
\begin{align}
    \label{eq:derivative_fidelity2_main}
     \left|\frac{d}{dt}\left|\braket{\psi_{E_{g}}(0)|\psi_{E_{g}}(t)}\right|\right|&\le|\braket{\psi_{E_{g}}(0)|(H-E_{g})|\psi_{E_{g}}(t)}|+\left|\braket{\psi_{E_{g}}(0)|\Gamma|\psi_{E_{g}}(t)}\right|.
\end{align}
We can bound the two terms on the right-hand side of Eq.~\eqref{eq:derivative_fidelity2_main} by physical quantities.
From Eq.~\eqref{eq:def_tilde_psi_main} and Eq.~\eqref{eq:def_psi_t_lambda}, the right-hand side can be decomposed as
\begin{align}
    &\braket{\psi_{E_{g}}(0)|(H-E_{g})|\psi_{E_{g}}(t)}=\sum_{l}p_{l}(0)\sum_{j}\epsilon_{j}e^{-i\left(\epsilon_{j}-i\gamma_{j}\right)t}|\braket{\epsilon_{j}|\psi_{l}(0)}|^{2}=\sum_{j}C_{j}\epsilon_{j}e^{-i\epsilon_{j}t-\gamma_{j}t},\label{eq:H_Eg_expectation1}\\&\braket{\psi_{E_{g}}(0)|\Gamma|\psi_{E_{g}}(t)}=\sum_{l}p_{l}(0)\sum_{j}\gamma_{j}e^{-i\left(\epsilon_{j}-i\gamma_{j}\right)t}|\braket{\epsilon_{j}|\psi_{l}(0)}|^{2}=\sum_{j}C_{j}\gamma_{j}e^{-i\epsilon_{j}t-\gamma_{j}t},
    \label{eq:Gamma_expectation1}
\end{align}
where $C_j\equiv \sum_l p_l(0) |\braket{\epsilon_j|\psi_l(0)}|^2\geq 0$. Therefore, we obtain
\begin{align}
    &\left|\braket{\psi_{E_{g}}(0)|(H-E_{g})|\psi_{E_{g}}(t)}\right|^{2}=\sum_{j,k}C_{j}C_{k}\epsilon_{j}\epsilon_{k}e^{-i\left(\epsilon_{j}-\epsilon_{k}\right)t}e^{-(\gamma_{j}+\gamma_{k})t}\nonumber\\ &=\sum_{j,k}C_{j}C_{k}\epsilon_{j}\epsilon_{k}\cos\left(\left(\epsilon_{j}-\epsilon_{k}\right)t\right)e^{-(\gamma_{j}+\gamma_{k})t}.
    \label{eq:H_Eg_expectation2}
\end{align}
From $\cos(x)\le 1$, $\epsilon_j\geq 0$ and $\gamma_j\geq 0$ for all $j$, we obtain the upper bound of the first term on the right-hand side in Eq.~\eqref{eq:derivative_fidelity2_main} as follows:
\begin{align}
    \left|\braket{\psi_{E_{g}}(0)|(H-E_{g})|\psi_{E_{g}}(t)}\right|&\le\left|\braket{\psi_{E_{g}}(0)|(H-E_{g})|\psi_{E_{g}}(0)}\right|=\braket{H}(0)-E_{g},
    \label{eq:equality_HS_main}
\end{align}
where we used $\ket{\psi_\lambda(0)}=\ket{\widetilde\psi(0)}=\ket{\psi(0)}$.
Similarly, we obtain the upper bound of the second term on the right-hand side in Eq.~\eqref{eq:derivative_fidelity2_main} as follows:
\begin{align}
    &\left|\braket{\psi_{E_{g}}(0)|\Gamma|\psi_{E_{g}}(t)}\right|^{2}=\sum_{j,k}C_{j}C_{k}\gamma_{j}\gamma_{k}\cos\left(\left(\epsilon_{j}-\epsilon_{k}\right)t\right)e^{-(\gamma_{j}+\gamma_{k})t}\nonumber\\ &\le\sum_{j,k}C_{j}C_{k}\gamma_{j}\gamma_{k}e^{-(\gamma_{j}+\gamma_{k})t}=\left(\sum_{j}C_{j}\gamma_{j}e^{-\gamma_{j}t}\right)^{2}.
    \label{eq:second_term_upperbound}
\end{align}
Substituting Eq.~\eqref{eq:equality_HS_main} and Eq.~\eqref{eq:second_term_upperbound} into Eq.~\eqref{eq:derivative_fidelity2_main}, we obtain
\begin{align}
    \label{eq:zero_t_UB_main}
    -\frac{d}{dt}|\braket{\psi_{E_{g}}(0)|\psi_{E_{g}}(t)}|&\le\left|\frac{d}{dt}|\braket{\psi_{E_{g}}(0)|\psi_{E_{g}}(t)}|\right|\le\braket{H}(0)-E_{g}-\frac{d}{dt}\sum_{j}C_{j}e^{-\gamma_{j}t}.
\end{align}
By integrating Eq.~\eqref{eq:zero_t_UB_main} from $t=0$ to $\tau$, Eq.~\eqref{eq:zero_t_UB_main} becomes
\begin{align}
    &1-|\braket{\psi_{E_{g}}(0)|\psi_{E_{g}}(\tau)}|\le\tau\left(\braket{H}(0)-E_{g}\right)+1-\sum_{j}C_{j}\exp\left(-\gamma_{i}\tau\right)\nonumber\\&\le\tau\left(\braket{H}(0)-E_{g}\right)+1-e^{-\tau\sum_{j}C_{j}\gamma_{j}}=\tau\left(\braket{H}(0)-E_{g}\right)+1-e^{-\braket{\Gamma}(0)\tau},
    \label{eq:integral_result_main}
\end{align}
where we used the Jensen's inequality for $\sum_j C_j=1$ in the second inequality and $\braket{\Gamma}(0)=\sum_j C_j \gamma_j$. From this inequality, we obtain Eq.~\eqref{eq:fidelity_main}.

\section{Derivation of Eq.~\eqref{eq:Lambda_MT_def}\label{sec:MT_fidelity_deriv}}
Let $A$ be an arbitrary Hermitian operator and $\ket{\phi}$ be a (normalized) state vector. 
Let $\ket{\phi_{\perp}}$ be a (normalized) vector orthogonal to $\ket{\phi}$. 
Then it is known that the following relation holds:
\begin{align}
    A\ket{\phi}=\langle A\rangle\ket{\phi}+\Delta A\ket{\phi_{\perp}}.
    \label{eq:A_decompose}
\end{align}
Equation~\eqref{eq:A_decompose} can be generalized to a non-Hermitian operator $\mathcal{H}(t) = H(t) - i \Gamma(t)$.
Since $\ket{\widetilde{\psi}(t)}$ is normalized by definition, we obtain
\begin{align}
    \label{eq:vec_expand_main}
    \alpha(t) \ket{\widetilde\psi_{\perp}(t)}=(\mathcal{H}(t)\otimes \mathbb{I}_A)\ket{\widetilde\psi(t)} -\langle \mathcal{H}(t)\rangle(t)\ket{\widetilde\psi(t)},
\end{align}
where $\ket{\widetilde\psi_{\perp}(t)}$ is a normalized state vector orthogonal to $\ket{\widetilde{\psi}(t)}$, i.e., $\braket{\widetilde\psi_{\perp}(t)|\widetilde\psi(t)}=0$ and $\alpha(t)$ is a coefficient to normalize $\ket{\widetilde\psi_{\perp}(t)}$.  
Taking the squared norm of this relation, we have 
\begin{align}
    \label{eq:def_alpha_main}
    |\alpha(t)|^2=\langle \mathcal{H}^\dagger(t) \mathcal{H}(t) \rangle(t)-|\langle \mathcal{H}(t)\rangle(t)|^2=\Delta \mathcal{H}(t)^2,
\end{align}
where 
\begin{align}
    \Delta O(t)\equiv\sqrt{\langle O^{\dagger}O\rangle(t)-|\langle O\rangle(t)|^{2}},
    \label{eq:Delta_O_def}
\end{align}
for an arbitrary operator $O$.
Equation~\eqref{eq:Delta_O_def} is a generalization of Eq.~\eqref{eq:stdev_def} for the arbitrary operator case. 
By differentiating $|\braket{\widetilde\psi(\tau_1)|\widetilde\psi(t)}|^2$, we obtain
\begin{align}
    &\frac{d}{dt}\left(|\braket{\widetilde{\psi}(\tau_{1})|\widetilde{\psi}(t)}|^{2}\right)=2\mathrm{Im}\left(\braket{\widetilde{\psi}(\tau_{1})|\mathcal{H}(t)|\widetilde{\psi}(t)}\braket{\widetilde{\psi}(t)|\widetilde{\psi}(\tau_{1}))}\right)-2\mathrm{Im}\langle\mathcal{H}(t)\rangle(t)|\braket{\widetilde{\psi}(\tau_{1})|\widetilde{\psi}(t)}|^{2}.
    \label{eq:timederiv_psitilde_fidelity}
\end{align}
Substituting Eq.~\eqref{eq:vec_expand_main} into Eq.~\eqref{eq:timederiv_psitilde_fidelity}, we have
\begin{align}
    \label{eq:diff_psi_alpha_main}
    &\left|\frac{d}{dt}\left(|\braket{\widetilde{\psi}(\tau_{1})|\widetilde{\psi}(t)}|^{2}\right)\right|=2\left|\mathrm{Im}\left(\alpha(t)\braket{\widetilde{\psi}(\tau_{1})|\widetilde{\psi}_{\perp}(t)}\braket{\widetilde{\psi}(t)|\widetilde{\psi}(\tau_{1})}\right)\right|\le2|\alpha(t)|\left|\braket{\widetilde{\psi}(\tau_{1})|\widetilde{\psi}_{\perp}(t)}\right|\left|\braket{\widetilde{\psi}(t)|\widetilde{\psi}(\tau_{1})}\right|.
\end{align}
Letting $\ket{\widetilde\psi_{\perp\perp}(t)}$ be a vector that is othogonal to both $\ket{\widetilde\psi(t)}$ and $\ket{\widetilde\psi_\perp(t)}$, we can expand as 
\begin{align}
    &\ket{\widetilde{\psi}(\tau_{1})}=\braket{\widetilde{\psi}(t)|\widetilde{\psi}(\tau_{1})}\ket{\widetilde{\psi}(t)}+\braket{\widetilde{\psi}_{\perp}(t)|\widetilde{\psi}(\tau_{1})}\ket{\widetilde{\psi}_{\perp}(t)}+\beta\ket{\widetilde{\psi}_{\perp\perp}(t)}.
    \label{eq:psi_tilde_expand}
\end{align}
From $\braket{\widetilde\psi(\tau_1)|\widetilde\psi(\tau_1)}=\braket{\widetilde\psi(t)|\widetilde\psi(t)}=1$, we have
\begin{align}
    \label{eq:perp_inequality_main}
    |\braket{\widetilde\psi_\perp(t)|\widetilde\psi(\tau_1)}|\le \sqrt{1- |\braket{\widetilde\psi(t)|\widetilde\psi(\tau_1)}|^2}.
\end{align}
Let $\cos\phi\equiv |\braket{\widetilde\psi(\tau_1)|\widetilde\psi(t)}|\le 1$ for $\phi\in[0,\pi / 2]$. Substituting this definition into Eq.~\eqref{eq:diff_psi_alpha_main} and using Eq.~\eqref{eq:perp_inequality_main} and $d\phi(t)/dt \le |d\phi(t)/dt|$, we have
\begin{align}
    \label{eq:diff_phi_main}
    \frac{d\phi}{dt}\le |\alpha(t)|=\Delta\mathcal{H}(t),
\end{align}
where we used Eq.~\eqref{eq:def_alpha_main}.
Equation~\eqref{eq:diff_phi_main} was derived in Ref.~\cite{PhysRevA.104.052620}. 
By integrating from $t=\tau_1$ to $\tau_2\geq\tau_1$, we obtain the speed limit:
\begin{align}
    \int_{\tau_{1}}^{\tau_{2}}dt\Delta\mathcal{H}(t)\geq\arccos\left(|\braket{\widetilde{\psi}(\tau_{1})|\widetilde{\psi}(\tau_{2})}|\right),
    \label{eq:LM_type_fidelity_main}
\end{align}
From this inequality, we obtain Eq.~\eqref{eq:Lambda_MT_def}.
\end{widetext}

\section{Classical Markov process\label{sec:classical_case}}

Consider the classical Markov process with $N$ states $\{B_1, B_2, \cdots, B_N\}$. Let $P(\nu, t)$ be the probability that the state equals $B_\nu$ at time $t$ and let $W_{\nu\mu}$ be the time-independent transition rate from state $B_\mu$ to state $B_\nu$. Taking $H_S=0$, $L_{\nu\mu}=\sqrt{W_{\nu\mu}}\ket{B_\nu}\bra{B_\mu}$ and $\rho_S(t) = \sum_\nu P(\nu, t)\ket{B_\nu}\bra{B_\nu}$, the Lindblad equation is reduced to the corresponding classical Markov process (see~\cite{Hasegawa:2023:BulkBoundaryBoundNC}). Since $[H_S, \sum_m L_m^\dagger L_m]=0$, we can apply the calculation in Subsection \ref{subsec:app_ML}.
 From Eq.~\eqref{eq:fidelity_open_main} and using the monotonicity of fidelity with respect to the partial trace, we obtain 
\begin{align}
    e^{-\frac{1}{2}\mathfrak{a}(0)\tau}&\le |\braket{\Psi(0)|\Psi(\tau)}|\le \sqrt{\mathrm{Fid}(\rho_{S}(0),\rho_{S}(\tau))}\nonumber\\
    &=\sum_{\nu} \sqrt{P(\nu,0) P(\nu, \tau)},
\end{align}
where where $\mathfrak{a}(0)\equiv\mathrm{Tr}\left[\sum_m L_m^\dagger L_m\rho_S(0)\right]=\sum_{\nu, \mu(\nu\neq\mu)} W_{\nu\mu}P(\mu,0)$.
This relation yields the speed limit: 
\begin{align}
    \tau\geq\frac{D_{\frac{1}{2}}\left(P(\nu, 0)\|P(\nu, \tau)\right)}{\mathfrak{a}(0)},
    \label{eq:ML_bound_classical}
\end{align}
where  $D_{1/2}(P\|Q)$ denotes the R\'{e}nyi divergence~\cite{TIT.2014.2320500} defined as 
\begin{align}
    D_\alpha(P\|Q) \equiv \frac{1}{\alpha-1}\ln\sum_{z \in\mathcal{Z}} P(z)^\alpha Q(z)^{1-\alpha}.
    \label{eq:D_alpha_def}
\end{align}
From Eq.~\eqref{eq:ML_type_UR_open_main}, we obtain the uncertainty relation:
\begin{align}
     &e^{\mathfrak{a}(0)\tau}-1\geq \left(\frac{\langle \hat{\mathcal{C}}\rangle(\tau)-\langle \hat{\mathcal{C}}\rangle(0)}{\Delta\hat{\mathcal{C}}(\tau)+\Delta\hat{\mathcal{C}}(0)}\right)^2,
     \label{eq:azero_tau_UR}
\end{align}
which agrees with Ref.~\cite{Hasegawa:2024:ConcentrationIneq}. 

\section{Models satisfying the commutative condition\label{sec:models}}

\begin{figure}
\includegraphics[width=1\linewidth]{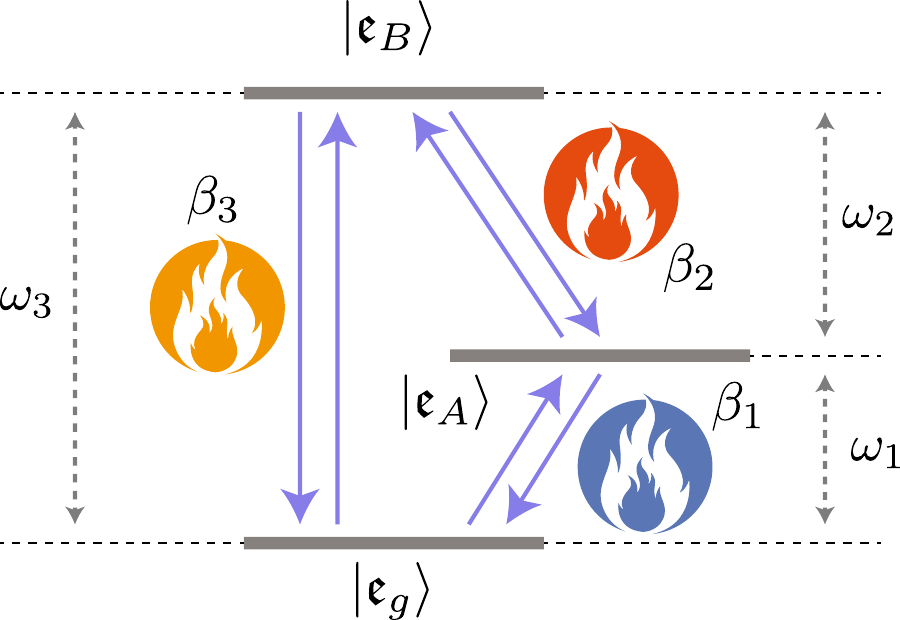}
    \caption{
Illustration of three states  autonomous refrigerator, which comprises three states: $\left|\mathfrak{e}_A\right\rangle$, $\left|\mathfrak{e}_B\right\rangle$, and $\left|\mathfrak{e}_g\right\rangle$. The transitions between these states are coupled with heat reservoirs, each characterized by  inverse temperature $\beta_r$ where $r=1,2,3$.    
    }
    \label{fig:quantum_heat_engine}
\end{figure}

In the case of Magolus-Levitin type tradeoff relations,
from Eq.~\eqref{eq:commutative_condition},
it is necessary that $H_S$ and $\sum_m L_m^\dagger L_m$ commute.
This condition is satisfied by the dephasing model and  autonomous refrigerator, which are reviewed in this section. 

\subsection{Dephasing model\label{sec:dephasing_model}}

The dephasing model serves to illustrate the reduction in coherence due to environmental interactions \cite{Nielsen:2011:QuantumInfoBook}. While this model is typically characterized as a CPTP, it can also be considered within the continuous time formalism. 
The Lindblad equation for the dephasing model is given by
\begin{align}
    \dot{\rho}_{S}&=\frac{\gamma}{2}\left(\sigma_{z}\rho_{S}\sigma_{z}-\rho_{S}\right)\nonumber\\&=\frac{\gamma}{2}\left(\sigma_{z}\rho_{S}\sigma_{z}-\frac{1}{2}\left\{ \sigma_{z}\sigma_{z},\rho_{S}\right\} \right),
    \label{eq:Lindblad_dephasing}
\end{align}
where $\sigma_z \equiv \ket{0}\bra{0} - \ket{1}\bra{1}$ is the Pauli-Z operator and $\gamma$ is the decay rate, and
we used the relation $\sigma_z^2 = \mathbb{I}_S$. 
Apparently, the Lindblad equation of Eq.~\eqref{eq:Lindblad_dephasing} satisfies the commutation condition $[H_S,L^\dagger L] = 0$.
When writing Eq.~\eqref{eq:Lindblad_dephasing} as the Kraus representation, we have
\begin{align}
    \rho_S(t+dt)=q\sigma_{z}\rho_S(t)\sigma_{z}+\left(1-q\right)\rho_S(t),
    \label{eq:dephasing_Kraus_repr}
\end{align}
where $q \equiv \gamma dt / 2$. Specifically, suppose that $\rho_{S}(t)$ is expressed as follows in the $\{\ket{0},\ket{1}\}$ basis:
\begin{align}
    \rho_S(t)=\left[\begin{array}{cc}
\rho_{00} & \rho_{01}\\
\rho_{10} & \rho_{11}
\end{array}\right].
\label{eq:rhot_expr}
\end{align}
Then, $\rho_{S}(t+dt)$ becomes 
\begin{align}
    \rho_S(t+dt)=\left[\begin{array}{cc}
\rho_{00} & (1-2q)\rho_{01}\\
(1-2q)\rho_{10} & \rho_{11}
\end{array}\right],
\label{eq:rhot_dt_expr}
\end{align}
which implies that the non-diagonal elements, i.e., coherence, decay due to interactions with the environment. 
This shows that, for a long time limit in Eq.~\eqref{eq:Lindblad_dephasing}, the initial coherence eventually fades away. 

\subsection{Autonomous refrigerator\label{sec:quantum_heat_engine}}

An  autonomous refrigerator  is an apparatus whose dynamics is described by quantum mechanics.
Refrigeration fundamentally works by moving heat from a cooler object to a warmer one. In a power refrigerator, this process performed by using mechanical power. An autonomous refrigerator operates using an additional work reservoir \cite{Mitchison:2019:QMachine}.

Consider a three-state quantum  autonomous refrigerator  as an example comprising the states $\ket{\mathfrak{e}_A}$, $\ket{\mathfrak{e}_B}$, and $\ket{\mathfrak{e}_g}$ \cite{Paule:2018:PhDBook}. This machine is powered by three heat reservoirs with different inverse temperatures $\beta_r$ ($r=1,2,3$). Each transition is coupled to each of the heat reservoirs. 
The Hamiltonian of this system can be represented as
\begin{align}
    H_S=\omega_{3}\ket{\mathfrak{e}_{B}}\bra{\mathfrak{e}_{B}}+\omega_{1}\ket{\mathfrak{e}_{A}}\bra{\mathfrak{e}_{A}},
    \label{eq:H_QHE_def}
\end{align}
where $\omega_1, \omega_2$, and $\omega_3=\omega_1+\omega_2$ denote the energy gaps between the levels (Fig.~\ref{fig:quantum_heat_engine}):
$\ket{\mathfrak{e}_{A}}\leftrightarrow\ket{\mathfrak{e}_{g}}$, $\ket{\mathfrak{e}_{B}}\leftrightarrow\ket{\mathfrak{e}_{A}}$, and $\ket{\mathfrak{e}_{B}}\leftrightarrow\ket{\mathfrak{e}_{g}}$.
The dynamics of the density operator follows the Lindblad equation [Eq.~\eqref{eq:Lindblad_eq_def}], where the jump operators are specified by
\begin{align}
    L_{\nu\mu} = \sqrt{W_{\nu\mu}}\ket{\mathfrak{e}_\nu}\bra{\mathfrak{e}_\mu},
    \label{eq:QHE_Lij_def}
\end{align}
where $\nu,\mu \in \{A,B,g\}$ and $W_{\nu\mu}$ is the transition rate defined by
\begin{align}
    W_{gA}&=\gamma\left(n_{1}^{\text{th }}+1\right)\label{eq:WgA},\\
    W_{Ag}&=\gamma n_{1}^{\text{th }}\label{eq:WAg},\\
    W_{AB}&=\gamma\left(n_{2}^{\text{th }}+1\right)\label{eq:WAB},\\
    W_{BA}&=\gamma n_{2}^{\text{th }}\label{eq:WBA},\\
    W_{gB}&=\gamma\left(n_{3}^{\text{th }}+1\right)\label{eq:WgB},\\
    W_{Bg}&=\gamma n_{3}^{\text{th }}\label{eq:WGb},
\end{align}
where
$n_r^{\text {th }} \equiv\left(e^{\beta_r \omega_r}-1\right)^{-1}$ and $\gamma$ is the decay rate. 
While the transitions induced by the jump operators are confined to those between energy eigenstates, making the dynamics similar to a classical Markov process, this is not always true. Specifically, if the initial density operator contains coherence, represented by off-diagonal terms with respect to the energy eigenstates, the dynamics deviate from the classical scenario.

\end{document}